%
%
%
\def\today{\ifcase\month\or January\or February\or March\or April\or May\or
June\or July\or August\or September\or October\or November\or December\fi
\space\number\day, \number\year}
%
%
\newcount\notenumber

\def\note{\global\advance\notenumber by 1 \footnote{$^{\the\notenumber}$}}
%
%
\newif\ifsectionnumbering
\newcount\eqnumber
\def\cleareqnumber{\eqnumber=0}
\def\numbereq{\global\advance\eqnumber by 1
\ifsectionnumbering\eqno(\the\secnumber.\the\eqnumber)\else\eqno
(\the\eqnumber)\fi}
\def\eqalinno{{\global\advance\eqnumber by 1}
\ifsectionnumbering(\the\secnumber.\the\eqnumber)\else(\the\eqnumber)\fi}
\def\name#1{\ifsectionnumbering\xdef#1{\the\secnumber.\the\eqnumber}
\else\xdef#1{\the\eqnumber}\fi}
\def\nosectionnumbering{\sectionnumberingfalse}
\sectionnumberingtrue
%
%
\newcount\refnumber

\immediate\openout1=refs.tex
\immediate\write1{\noexpand\frenchspacing}
\immediate\write1{\parskip=0pt}
\def\ref#1#2{\global\advance\refnumber by 1%
[\the\refnumber]\xdef#1{\the\refnumber}%
\immediate\write1{\noexpand\item{[#1]}#2}}
\def\refx#1#2{\global\advance\refnumber by 1%
{\the\refnumber}\xdef#1{\the\refnumber}%
\immediate\write1{\noexpand\item{[#1]}#2}}
%
\def\tie{\noexpand~}

%
%
\font\twelvebf=cmbx10 scaled \magstep1
\newcount\secnumber

\def\newsection#1.{\ifsectionnumbering\cleareqnumber\else\fi%
        \global\advance\secnumber by 1%
        \bigbreak\bigskip\par%
        \line{\twelvebf \the\secnumber. #1.\hfil}\nobreak\medskip\par\noindent}
%
%
%
\def \sqr#1#2{{\vcenter{\vbox{\hrule height.#2pt
        \hbox{\vrule width.#2pt height#1pt \kern#1pt
                \vrule width.#2pt}
                \hrule height.#2pt}}}}
\def\Box{{\mathchoice\sqr54\sqr54\sqr33\sqr23}\,}
%
%
%
\newdimen\fullhsize
\def\fiddle{\fullhsize=6.5truein \hsize=3.2truein}
\def\fullline{\hbox to\fullhsize}
\def\mkhdline{\vbox to 0pt{\vskip-22.5pt
        \fullline{\vbox to8.5pt{}\the\headline}\vss}\nointerlineskip}
\def\mkftline{\baselineskip=24pt\fullline{\the\footline}}
\let\lr=L \newbox\leftcolumn
\def\twocolumns{\fiddle
        \output={\if L\lr \global\setbox\leftcolumn=\columnbox
                \global\let\lr=R \else \doubleformat \global\let\lr=L\fi
                \ifnum\outputpenalty>-20000 \else\dosupereject\fi}}
\def\doubleformat{\shipout\vbox{\mkhdline
                \fullline{\box\leftcolumn\hfil\columnbox}
                \mkftline} \advancepageno}
\def\columnbox{\leftline{\pagebody}}
\nosectionnumbering
\magnification=1200
\def\pr#1 {Phys. Rev. {\bf D#1\tie }}
\def\pe#1 {Phys. Rev. {\bf #1\tie}}
\def\pre#1 {Phys. Rep. {\bf #1\tie}}
\def\pl#1 {Phys. Lett. {\bf #1B\tie }}
\def\prl#1 {Phys. Rev. Lett. {\bf #1\tie }}
\def\np#1 {Nucl. Phys. {\bf B#1\tie }}
\def\ap#1 {Ann. Phys. (NY) {\bf #1\tie }}
\def\cmp#1 {Commun. Math. Phys. {\bf #1\tie }}
\def\imp#1 {Int. Jour. Mod. Phys. {\bf A#1\tie }}
\def\mpl#1 {Mod. Phys. Lett. {\bf A#1\tie}}
\def\spn#1 {Sov. Phys. JETP {\bf #1\tie}}
\def\jetp#1 {J. Exp. Theor. Phys. {\bf #1\tie}}
\def\tie{\noexpand~}

\parskip=15pt plus 4pt minus 3pt
\headline{\ifnum \pageno>1\it\hfil Supersymmetry and
Gravitational   $\ldots$\else \hfil\fi}
\font\title=cmbx10 scaled\magstep1
\font\tit=cmti10 scaled\magstep1
\footline{\ifnum \pageno>1 \hfil \folio \hfil \else
\hfil\fi}
\raggedbottom


\def\half{{\textstyle{1\over2}}}
\def\ha{{1\over2}}
\overfullrule0pt


\rightline{\vbox{\hbox{RU99-07-B}\hbox{NYU-TH/99/07/01}\hbox
{CERN-TH/99-195}\hbox{hep-th/9909012}}}
\vfill
\centerline{\title SUPERSYMMETRY AND GRAVITATIONAL QUADRUPOLES}
\vfill
{\centerline{\title Ioannis Giannakis${}^{a}$,
James T.~Liu${}^{a,d}$ and Massimo Porrati${}^{a, b, c}$ \footnote{$^{\dag}$}
{\rm e-mail: \vtop{\baselineskip12pt
\hbox{giannak@theory.rockefeller.edu, jimliu@umich.edu,}
\hbox{massimo.porrati@nyu.edu}}}}
}
\medskip
\centerline{$^{(a)}${\tit Physics Department, The Rockefeller
University}}
\centerline{\tit 1230 York Avenue, New York, NY
10021-6399}
\medskip
\centerline{$^{(b)}${\tit Department of Physics, New York University}}
\centerline{\tit 4 Washington Pl., New York, NY 10003}
\medskip
\centerline{$^{(c)}${\tit Theory Division, CERN}}
\centerline{\tit CH 1211 Geneva 23 (Switzerland)}
\medskip
\centerline{$^{(d)}${\tit Randall Laboratory of Physics,
University of Michigan}}
\centerline{\tit Ann Arbor, MI 48109}
\vfill
\centerline{\title Abstract}
\bigskip
{\narrower\narrower
We derive model independent, non-perturbative supersymmetric sum rules
for the gravitational quadrupole moments of arbitrary-spin particles in
any $N=1$ supersymmetric theory.  These sum rules select a
``preferred'' value of $h=1$ where the ``$h$-factor'' is the gravitational
quadrupole analog of the gyromagnetic ratio or $g$-factor.  This value
of $h=1$ corresponds identically to the preferred field theory value
obtained by tree-level unitarity considerations.  The presently derived
$h$-factor sum rule complements and generalizes previous work on
electromagnetic moments where $g=2$ was shown to be preferred by both
supersymmetric sum rule and tree-level unitarity arguments.

\par}
\vfill\vfill\break


\newsection Introduction.%
In models with unbroken supersymmetry, it is often the case that the nature
of the supersymmetry algebra leads to many powerful results on the structure
of the theory itself.  Such results often take the form of model independent
sum rules that must be obeyed by the states in the supersymmetric spectrum.
For example, the added constraint of unbroken supersymmetry requires that the
gyromagnetic ratio of the electron (as the spin-1/2 member of a chiral
multiplet) is given exactly by $g=2$
\ref\fr{S. Ferrara and E. Remiddi, Phys. Lett. {\bf 53B} (1974) 347.}.
Such results have since been generalized to yield constraints on the
electromagnetic multipole moments of any massive particle state in $N=1$
[%
\refx\por{S. Ferrara and M. Porrati, Phys. Lett. {\bf B288} (1992) 85.},%
\refx\glp{I. Giannakis, J. T. Liu and M. Porrati, \pr58 (1998) 045016, {\tt
hep-th/9803073}.}]
and $N=2$ 
\ref\gia{I. Giannakis and J. T. Liu, \pr58 (1998) 025009, {\tt
hep-th/9711173}.}
supermultiplets.  These sum rules were derived using general principles and
hold for any electromagnetic current commuting with supersymmetry.

In this letter we apply similar techniques to the study of gravitational
interactions in a theory with unbroken supersymmetry and derive sum rules
for the gravitational quadrupole moments of the particle states in arbitrary
$N=1$ massive supermultiplets.  Following
\ref\giamn{I. Giannakis, J. T. Liu and M. Porrati, \pr59 (1999) 104013, {\tt
hep-th/9809142}.},
we define an ``$h$-factor'', which is the gravitational analog of the
gyromagnetic ratio.  The $h$-factors of the different states in a
massive $N=1$ supermultiplet are no longer arbitrary, but are related
through the gravitational quadrupole sum rule; they are determined
in terms of a single real quantity (which may be taken as the $h$-factor
of the central member of the supermultiplet).  Furthermore, supersymmetry
selects a ``natural'' value of $h=1$, which in addition has the property of
yielding tree-level unitarity up to the Planck scale for gravitational
interactions
[%
\refx\porg{M. Porrati, Phys. Lett. {\bf B304} (1993) 77.},%
\refx\cpd{A. Cucchieri, M. Porrati and S. Deser, Phys. Rev. {\bf D51} (1995)
4543.},%
\giamn].
In this sense, the $h$-factor sum rule and its consequences resemble
those obtained for the usual gyromagnetic ratio, where the preferred
value of $g=2$ was similarly selected by both tree-level unitarity
[%
\refx\wein{S. Weinberg in {\it Lectures on Elementary Particles and
Quantum Field Theory}, S. Deser, M. Grisaru and H. Pendleton, eds.,
MIT Press, Cambridge, MA (1970).},%
\refx\fpt{S. Ferrara, M. Porrati and V.L. Telegdi, Phys. Rev. {\bf D46}
(1992) 3529.}]
and supersymmetric sum rule [\por] considerations (for yet another
reason, see
\ref\jac{R. Jackiw, Phys. Rev. {\bf D57} (1998) 2635.}).

Indeed our derivation of the $h$-factor sum rules follows the method
of [\por], and involves the transformation properties of a conserved
stress tensor $T_{\mu\nu}$ that commutes with the $N=1$ supersymmetry
algebra.  The main complication in obtaining the sum rules, that was not
present in [\por], is the requirement of working to second order
in the momentum transfer in the appropriate matrix elements.  This,
however, has been addressed in [\glp], where a generalization to
arbitrary electromagnetic multipole moments was presented.

Although the procedure used in obtaining the sum rules is by now standard,
care must be taken in manipulating the supersymmetry algebra since
various factors from different sources have to be combined in a precise
manner.  We follow the conventions for the Dirac algebra and $N=1$
supersymmetry given in [\glp].  Thus we start with the $N=1$ algebra
written in the form $\lbrace Q_{\alpha}, {\overline Q}_{\beta} \rbrace
= 2({\gamma}^{\mu})_{\alpha\beta} P_{\mu}$, so that in the rest frame of
a massive single particle state, we have
$$
\lbrace Q^{L}_{1\over 2}, Q^{R}_{-{1\over 2}} \rbrace=2M,
\qquad
\lbrace Q^{L}_{-{1\over 2}}, Q^{R}_{1\over 2} \rbrace=2M.
\numbereq\name{\eqana}
$$
Note that the superscripts $L,R$ denote chiralities, 
$$
{\gamma_5}Q^{L\atop R}=\pm Q^{L\atop R},
\numbereq
$$
whereas the subscripts $\pm\ha$ denote helicities,
$$
i\gamma^{12}Q_{\pm{1\over 2}}=\pm Q_{\pm {1\over 2}}.
\numbereq
$$
After proper normalization, the supersymmetry algebra, (\eqana),
corresponds to a Clifford algebra with two fermionic degrees of freedom.
Thus irreducible massive $N=1$ representations are generated by acting
with creation operators $Q^R_{\pm\ha}$ on a superspin-$j$ Clifford
vacuum, $|j\rangle$, that is annihilated by $Q^L_{\pm\ha}$.  As a
result, this generates the massive $N=1$ superspin-$j$ multiplet with
spins $(j+\ha,j,j,j-\ha)$.

Since the supercharges $Q^{L, R}_{\pm{1\over 2}}$
are operators of spin $1/2$, we use a shorthand notation of labeling the
spin-$j$ Clifford vacuum by $|0\rangle$, along with the superpartner
states $|{\uparrow}\rangle$, $|{\downarrow}\rangle$ and
$|{\updownarrow}\rangle$, which are generated by (normalized)
$Q^R_\ha$, $Q^R_{-\ha}$ and $Q^R_\ha Q^R_{-\ha}$, respectively.  While
this notation is initially convenient, it only addresses the action of
the supercharges on the Clifford vacuum.  The actual states in the
multiplet are only obtained after taking the Clebsch-Gordan combination
of angular momenta between the spin-$j$ Clifford vacuum and the spins
$[(1/2) + 2(0)]$, corresponding to
$[(|{\uparrow}\rangle,|{\downarrow}\rangle)+ |0\rangle+
|{\updownarrow}\rangle]$.

Since the gravitational quadrupole moment is defined in terms of moments
of the stress tensor, we proceed with an examination of the multiplet
containing $T_{\mu\nu}$.  For a superconformal theory, this is the
multiplet of currents $(T_{\mu\nu}, j_{\mu\alpha}, j_{\mu}^{(5)})$,
where $\alpha$ is a spinor index.  More generally, theories invariant
under $N=1$ supersymmetry may nevertheless violate dilatation
invariance, $T_{\mu}^{\mu} \ne 0$, supersymmetric invariance,
${\gamma^{\mu}}j_{\mu} \ne 0$, or $R$-invariance, ${\partial^{\mu}}
 j_{\mu}^{(5)} \ne 0$, or any combination of these symmetries.

We first consider the case of ordinary $N=1$ supersymmetry, without
conformal invariance, but with a conserved $R$-current:
$$
T_{\mu}^{\mu} \ne 0, \qquad {\gamma^{\mu}}j_{\mu} \ne 0, \qquad
{\partial^{\mu}}j_{\mu}^{(5)}=0.
\numbereq\name{\eqavon}
$$
The resulting multiplet of currents contains ($j_{\mu}^{(5)}, j_{\mu},
T_{\mu\nu}, t_{\mu\nu}$), where $t_{\mu\nu}$ is a conserved
antisymmetric tensor, and has the following transformation properties:
$$
\eqalign{
{\delta}j_{\mu}^{(5)}&=i {\overline{\epsilon}}{\gamma_5}j_{\mu},\cr
{\delta}j_{\mu}&=[-2i{\gamma^{\nu}}(T_{\mu\nu}+t_{\mu\nu})
-{\gamma_5} {\gamma^{\lambda}} {\partial_{\lambda}}j_{\mu}^{(5)}
-{i\over 2}{\epsilon_{\mu\nu\lambda\sigma}}
{\gamma^{\nu}}{\partial^{\lambda}}j^{\sigma (5)}] {\epsilon},\cr
{\delta}T_{\mu\nu}&=-{1\over 2}{\overline{\epsilon}}{\gamma_{(\mu}{}^{\lambda}}
{\partial_{\lambda}}j_{\nu)},\cr
{\delta}t_{\mu\nu}&=-{i\over 4}{\epsilon_{\mu\nu\lambda\sigma}}
{\overline{\epsilon}} {\gamma_5}{\gamma^{\sigma}}{\gamma^{\rho}}
{\partial^{\lambda}}j_{\rho}.\cr}
\numbereq\name{\eqonic}
$$
We find that
$[{\delta}_1, {\delta}_2]O=-2i({\overline{\epsilon}}_2{\gamma}^{\lambda}
{\epsilon}_1){\partial_\lambda}O$ where $O=(j_{\mu}^{(5)}, j_{\mu},
T_{\mu\nu}, t_{\mu\nu})$, matching our normalization of the $N=1$
algebra.
It follows then that two successive supersymmetry transformations on the
conserved current, $T_{\mu\nu}$, gives
$$
\eqalign{
{\delta}_{\eta}{\delta}_{\epsilon}T_{\mu\nu}=&
\,i{\overline{\epsilon}}\gamma_{(\mu}{}^{\lambda\rho}\eta
{\partial_\lambda}(T_{\nu)\rho}+t_{\nu)\rho})
-i{\overline{\epsilon}}{\gamma^{\lambda}}\eta
{\partial_\lambda}T_{\mu\nu}\cr
&+{3\over 4}{\overline{\epsilon}}{\gamma_5} \gamma^\lambda\eta
(\Box \eta_{\lambda(\mu}^{\vphantom{(5)}}
-\partial_{\lambda\vphantom{)}}^{\vphantom{(5)}}
\partial_{(\mu}^{\vphantom{(5)}})
j_{\nu)}^{(5)}
-{1\over 4}{\overline{\epsilon}}{\gamma_5}{\gamma^{\lambda}}\eta
(\Box\eta_{\mu\nu}-\partial_\mu\partial_\nu) j_{\lambda}^{(5)}.\cr}
\numbereq\name{\eqpasa}
$$
The matrix elements of this equation between single particle states
which belong to the same $N=1$ multiplet give rise to sum rules for
the gravitational multipoles of the particle states.

The gravitational moments may be obtained by performing a multipole
expansion of the matrix elements $\langle j',m',\vec p\,|T_{\mu\nu}
|j,m,0\rangle$ of the stress tensor.  While rotational symmetry allows
the matrix elements to be decomposed in terms of spherical tensors
(see {\it e.g.}
\ref\ren{V. Rahal and H. C. Ren, \pr41 (1989) 1989.})
or cartesian tensors of definite angular momentum
\ref\kst{K.S. Thorne, Rev. Mod. Phys. {\bf 52} (1980) 299.}, we find it
sufficient to take a general expansion with indefinite angular momentum
terms:
$$
\eqalign{
\langle j', m', \vec p\,|T_{00}|j, m, 0\rangle&=
\sum_{l=0}^{\infty}{1\over {l!}}(ip)_{i_1}(ip)_{i_2} \cdots
(ip)_{i_l}\langle j', m', 0,|M_{{i_1}{i_2} \cdots {i_l}}^{(l)}
|j, m, 0\rangle\cr
&=\langle 0\,|M^{(0)}|0\rangle +ip^i
\langle 0\,|M^{(1)}_{i}|0\rangle-{1\over 2!}p^{i}p^{j}
\langle 0\,|M^{(2)}_{ij}|0\rangle+ \cdots, \cr
\langle j', m', \vec p\,|T_{0i}|j, m, 0\rangle&=
\sum_{l=0}^{\infty}{1\over {(l+1)!}}
(ip)_{i_1}(ip)_{i_2} \cdots
(ip)_{i_l}\langle j', m', 0,|N_{i\,{i_1} \cdots {i_l}}^{(l+1)}
|j, m, 0\rangle \cr
&=\langle 0\,|N^{(1)}_{i}|0\rangle+{i\over {2!}}p^{j}
\langle 0\,|N^{(2)}_{i\,j}|0\rangle -{1\over {3!}}p^{j}p^{k}
\langle 0\,|N^{(3)}_{i\,jk}|0\rangle + \cdots, \cr
\langle j', m', \vec p\,|T_{ij}|j, m, 0\rangle&=
\sum_{l=0}^{\infty}{1\over {(l+2)!}}
(ip)_{i_1}(ip)_{i_2} \cdots
(ip)_{i_l}\langle j', m', 0,|L_{ij\,{i_1} \cdots {i_l}}^{(l+2)}
|j, m, 0\rangle\cr
&={1\over 2!}\langle 0\,|L^{(2)}_{ij}|0\rangle+{i\over{3!}}p^{k}
\langle 0\,|L^{(3)}_{ij\,k}|0\rangle-{1\over {4!}}
p^{k}p^{l}
\langle 0\,|L^{(4)}_{ij\,kl}|0\rangle + \cdots, \cr}
\numbereq\name{\eqnini}
$$
where some of the quantum numbers $(j',m')$ and $(j,m)$ have been
suppressed.  Note that the first index of $N^{(l+1)}$ and the first two
indices of $L^{(l+2)}$ are distinct from the remaining ones, and do not
share any definite symmetry properties.

To $p^2$ order, conservation of the stress tensor, $p^{\mu}T_{\mu\nu}=0$,
provides us with the following relations:
$$
\eqalign{
&M^{(0)}+iM{{p^{i}p^{j}}\over {p^2}}N^{(2)}_{i\,j}=0, \qquad N^{(1)}_{i}=0,
\qquad ip^{i}[M^{(1)}_{i}-{i\over 2M}N^{(1)}_{i}+{{iM}\over
3}{{p^{j}p^{k}}\over {p^2}}
N^{(3)}_{i\,jk}]=0, \cr
&{1\over 2}p^{i}p^{j}[M^{(2)}_{ij}-{i\over 2M}N^{(2)}_{i\,j}
+{{iM}\over 6}{{p^{k}p^{l}}\over {p^2}}
N^{(4)}_{i\,jkl}]=0, \qquad N^{(1)}_{i}
+{iM\over 3}{{p^{j}p^{k}}\over {p^2}}L^{(3)}_{ij\,k}=0, \cr
&p^{i}[iN^{(2)}_{ji}+{1\over 2M}L^{(2)}_{ij}-{M\over 6}{{p^{k}p^{l}}\over
{p^2}}
L^{(4)}_{ij\,kl}]=0, \qquad p^{j}L^{(2)}_{ij}=0. \cr}
\numbereq\name{\eqnonig}
$$
By multiplying the first and penultimate equations by $p^2$,
and by differentiating with respect to the momentum, we find the
relations
$$
N^{(2)}_{(i\,j)_S}=i{M^{(0)}\over M}\delta_{ij},\qquad  L^{(4)}_{i(j\,kl)_S}=
-2{M^{(0)}\over M^2}(\delta_{ij}\delta_{kl} + \delta_{ik}\delta_{jl} +
\delta_{il}\delta_{jk}),
\numbereq\name{\eqmpnb}
$$
where $()_S$ denotes complete symmetrization with weight one.

Note that
the gravitational dipole term in the expansion of $T_{00}$, $M^{(1)}_{i}$,
can be removed by translating $T_{00}$ and choosing the translation
$d_i$ appropriately, {\it i.e.}:
$$
\eqalign{
\langle \vec p\,| T_{00}(\vec x+\vec d)|0\rangle&=
\langle \vec p\,| \exp(id_ip^i)T_{00}(x)|0\rangle =
\langle \vec p\,|(1+id_ip^i+..)T_{00}|0\rangle\cr
&=\langle \vec p\,| T_{00}|0\rangle + iM^{(0)} d_ip^i + O(p^2)
=M^{(0)}+ip^{i}M^{(1)}_i +iM^{(0)} d_ip^i+\cdots. \cr}
\numbereq\name{\eqnigiu}
$$
Therefore there is no physical content in $M^{(1)}_i$, and we take it to
vanish henceforth.

It was demonstrated in [\giamn] that, outside of the Newtonian limit,
the nonrelativistic definition of the gravitational quadrupole has to be
amended.  The general expression for the quadrupole moment (in $D=4$)
is [\giamn]:
$$
Q^{ij}=\int d^3x\left(x^ix^j -{\textstyle{1\over 3}}\delta^{ij}x^2\right)
{\cal T}_{00},
\numbereq\name{\eqafhdig}
$$
where ${\cal T}_{00}= T_{00}+{\delta^{ij}}T_{ij}$.
By writing ${\cal T}_{00}(x)= \int {d^{3}p\over (2{\pi})^3}
{\overline {\cal T}}_{00}(p)e^{-ipx}$, we find that
$$
\eqalign{
Q_{ij}&=\left.\bigl[(i{{\partial}\over {{\partial}p^{i}}})(i{{\partial}\over
{{\partial}p^{j}}})
-{1\over 3}{\delta_{ij}}(i{{\partial}\over {{\partial}p^{k}}})(i{{\partial}
\over {{\partial}p_{k}}})\bigr]
{\overline {\cal T}}_{00}\right|_{p=0} \cr
&=M^{(2)}_{ij}-{1\over 3}{\delta_{ij}}M^{(2)k}_{k}+{1\over
12}{\delta^{kl}}L^{(4)}_{ij\,kl}
-{1\over {36}}{\delta_{ij}}{\delta^{ab}}{\delta^{kl}}L^{(4)}_{ab\,kl}.\cr}
\numbereq\name{\equabco}
$$
Similarly the definitions for the mass $M=\int d^3x\,T_{00}$ 
and the angular momentum $J_{i}=\int d^3x\,{\epsilon_{ijk}}
x^{j}T^{0k}$ provide us
with the relations $M=M^{(0)}$ and $J_{i}=-
{1\over 2}{\epsilon_{ijk}}N^{(2)j\,k}$.
The relevant expansions for the matrix elements of ${\cal T}_{00}$ and
$T_{0i}$ then become
$$
\eqalign{
\langle j', m', \vec p\,|{\cal T}_{00}|j, m, 0\rangle&=
M_j{\delta_{jj'}}{\delta_{mm'}}-{1\over 2}p^{i}p^{j}
\langle j', m', 0\,|Q_{ij}|j, m, 0\rangle+ \cdots, \cr
\langle j', m', \vec p\,|T_{0i}|j, m, 0\rangle&=
{i\over 2}p^{j}{\delta_{jj'}}{\delta_{mm'}}+{i\over 2}{\epsilon_{ijk}}p^{j}
\langle j', m', 0\,|J^{k}|j, m, 0\rangle +\cdots, \cr}
\numbereq\name{\equanvo}
$$
where we have omitted trace terms ({\it i.e.}~terms proportional to the
contracted momentum $p^2$).

As in the electromagnetic case [\glp], general multipole moment sum rules
are derived by taking the double supersymmetry variation of the conserved
stress tensor $T_{\mu\nu}$,
$$
\eqalign{
\delta_\eta\delta_\epsilon T_{\mu\nu} &=
[\overline{\eta} Q, [\overline{\epsilon}Q, T_{\mu\nu} ]]\cr
&=\overline{\eta}Q\overline{\epsilon}QT_{\mu\nu}
-\overline{\eta}QT_{\mu\nu} \overline{\epsilon}Q
-\overline{\epsilon}QT_{\mu\nu} \overline{\eta}Q
+T_{\mu\nu} \overline{\epsilon}Q\overline{\eta}Q,\cr}
\numbereq\name{\eqsamba}
$$
and evaluating it between single particle states $\langle\alpha|$ and
$|\beta\rangle$.  Since $\delta_\eta\delta_\epsilon T_{\mu\nu}$ is known,
and is given by (\eqpasa), the right hand side of (\eqsamba) then
relates the matrix elements of $T_{\mu\nu}$ among the superpartner
states generated by $Q$.  By choosing the initial and final states to be
the superspin-$j$ Clifford vacuum, and by picking $\eta_L=0$, we find
$$
\langle\alpha, \vec p\,|\delta_{\eta_R}\delta_\epsilon T_{\mu\nu}
|\beta, 0\rangle
= \langle\alpha,\vec p\,|T_{\mu\nu}\overline{\epsilon}Q\overline{\eta}_RQ
|\beta,0\rangle
- \langle\alpha,0|\overline{\epsilon}Q^{(p)}L^{-1}(\vec p\,)T_{\mu\nu}
\overline{\eta}_RQ|\beta,0\rangle,
\numbereq\name{\gensum}
$$
where $Q^{(p)}$ denotes the Lorentz boost of $Q$, namely
$Q^{(p)}=L^{-1}(\vec p\,)QL(\vec p\,)$,
and $|\alpha, \vec p\ \rangle=L(\vec p\,)|\alpha, 0\rangle$.  This
expression provides the basis of determining both the ``vanishing'' and
``diagonal'' sum rules.

For the vanishing sum rule, we set $\epsilon_L=0$ so that only
$\epsilon_R$ is active.  Since Eq.~(\eqpasa) indicates that
$\delta_{\eta_R}\delta_{\epsilon_R}T_{\mu\nu}=0$, the sum rule
(\gensum) becomes
$$
\langle\alpha,\vec p\,|T_{\mu\nu}\overline{\epsilon}_RQ\overline{\eta}_RQ
|\beta,0\rangle = 0.
\numbereq
$$
This demonstrates that {\it all} matrix elements of the stress tensor
vanish between states $|0\rangle$ and $\left|\updownarrow\right\rangle$,
and hence that there are no off-diagonal moments between the two spin-$j$
states of the supermultiplet.

For the diagonal sum rule, we choose $\epsilon_R=0$ and find
instead
$$
\eqalign{
\langle\alpha,0|
\overline{\epsilon}_LQL^{-1}(\vec p\,)T_{\mu\nu}
\overline{\eta}_R&Q|\beta,0\rangle =
2M(\overline{\epsilon}_L{\gamma^0}{\eta_R})\langle\alpha, \vec p\,|T_{\mu\nu}
|\beta, 0\rangle\cr
&-\langle\alpha, \vec p\,|\delta_{\eta_R}\delta_{\epsilon_L} T_{\mu\nu}
|\beta, 0\rangle
-\langle\alpha,0|\lbrack L^{-1}(\vec p\,),
\overline{\epsilon}_LQ
\rbrack T_{\mu\nu}
\overline{\eta}_RQ|\beta,0\rangle.\cr}
\numbereq\name{\genbcvx}
$$
The first line of (\genbcvx) essentially relates the matrix elements in
the superpartner state with those in the original state, while the
second line provides correction terms originating from supersymmetry and
from Lorentz boosts.

For the Lorentz boost correction, we make use of the fact that $Q$
transforms as a spinor to derive
$$
\eqalign{
\lbrack L^{-1}(\vec p\,), \overline{\epsilon}_LQ
\rbrack& =\overline{\epsilon}_L \left(
{\sqrt{{E+M}\over 2M}}\bigl(1+{p^{i}\over {(E+M)}}{\gamma^{0i}}\bigr)-
1\right)QL^{-1}(\vec p\,) \cr
&={p^{i}\over 2M}\overline{\epsilon}_L{\gamma^{0i}}QL^{-1}(\vec p\,)
+\cdots, \cr}
\numbereq\name{\eqnvncxus}
$$
where the terms that have been dropped are all trace-like, and do not
contribute to the static moments.  Specializing to the gravitational
quadrupole moment, we are interested in the supersymmetry variation
$\delta_{\eta_R}\delta_{\epsilon_L}{\cal T}_{00}$.  Using
Eq.~(\eqpasa) we find that
$$
\eqalign{
{\delta}_{\eta_R}{\delta}_{\epsilon_L}T_{00}=&
-p_{\lambda}{\overline{\epsilon}}_L\gamma_{0}{}^{\lambda\rho}\eta_R
(T_{0\rho}+t_{0\rho})+p_{\lambda}{\overline{\epsilon}}_L
\gamma^{\lambda}\eta_R
T_{00}\cr
&-{3\over 4}{\overline{\epsilon}}_L{\gamma_5}
{\gamma^{\lambda}}\eta_R ({\eta_{0\lambda}}p^2-p_0p_{\lambda}) j_{0}^{(5)}
+{1\over 4}{\overline{\epsilon}}_L{\gamma_5}
{\gamma^{\lambda}}\eta_R ({\eta_{00}}p^2-p_0p_0) j_{\lambda}^{(5)}\cr
=&-p_{i}({\overline{\epsilon}}_L\gamma_{0}{}^{ij}\eta_R)
(T_{0j}+t_{0j})+p_{i}({\overline{\epsilon}}_L\gamma^{i}\eta_R)T_{00}
+O(p^2) \cr}
\numbereq\name{\eqpasanb}
$$
and
$$
\eqalign{
{\delta^{ij}}{\delta}_{\eta_R}{\delta}_{\epsilon_L}T_{ij}=&
-p_{\lambda}{\overline{\epsilon}}_L\gamma_{0}{}^{\lambda\rho}\eta_R
(T_{\rho0}+t_{\rho0})+p_{\lambda}{\overline{\epsilon}}_L\gamma^{\lambda}\eta_R
{\delta^{ij}}T_{ij}+p_{k}{\overline{\epsilon}}_L\gamma^{ikj}\eta_R
t_{ij} \cr
&-{3\over 4}{\overline{\epsilon}}_L{\gamma_5}
{\gamma^{\lambda}}\eta_R (\eta_{i\lambda} p^2 -p_ip_\lambda)j_i^{(5)}
+{1\over 4}{\overline{\epsilon}}_L{\gamma_5}
{\gamma^{\lambda}}\eta_R ({\eta_{ii}}p^2-p_ip_i) j_{\lambda}^{(5)}\cr
=&-p_{i}({\overline{\epsilon}}_L\gamma_{0}{}^{ij}\eta_R)
(T_{j0}+t_{j0})+p_{k}
({\overline{\epsilon}}_L\gamma^{k}\eta_R){\delta^{ij}}T_{ij}
+O(p^2). \cr}
\numbereq\name{\eqpasanbqy}
$$
Note that the antisymmetric tensor $t_{\mu\nu}$ drops out when the two
equations are combined:
$$
\delta_{\eta_R}\delta_{\epsilon_L}{\cal T}_{00}=
-2p_i(\overline{\epsilon}_L\gamma_0{}^{ij}\eta_R)T_{0j}
+p_i(\overline{\epsilon}_L\gamma^i\eta_R){\cal T}_{00}+O(p^2).
\numbereq\name{\eqddct}
$$
In particular, this is a consequence of the complete relativistically
correct definition of the gravitational quadrupole moment, (\eqafhdig).

The diagonal sum rule now takes the form:
$$
\eqalign{
{1\over 2M} \langle\alpha,0|\overline{\epsilon}_LQL^{-1}(\vec p\,)
{\cal T}_{00}\overline{\eta}_RQ|\beta,0\rangle=&({\overline{\epsilon}}_L
\gamma^{0}\eta_R)\langle\alpha, \vec p\, |{\cal T}_{00}|\beta,0\rangle \cr
&-i{\epsilon^{ijk}}{{p_j}\over
{M}}({\overline{\epsilon}}_L \gamma_k\eta_R)
\langle\alpha, \vec p\, | T_{0i}|\beta,0\rangle, \cr}
\numbereq\name{\eqpasad}
$$
where we have omitted terms of order $p^2$.  Due to rotational
invariance it is sufficient to examine the 33 component of the
quadrupole moment.  By substituting the
definitions of the quadrupole and angular momentum matrix elements,
(\equanvo), into Eq.~(\eqpasad) and by equating terms of the same order
in momentum transfer, we arrive at the following diagonal sum rules:
$$
\eqalign{
\langle{\uparrow}|Q_{33}|{\uparrow}\rangle&=\langle0|Q_{33}|0\rangle
+{2\over {3M}}\langle0|J_{3}|0\rangle, \qquad
\langle{\uparrow}|Q_{33}|{\downarrow}\rangle=
-{1\over3M}\langle0|J_-|0\rangle,\cr
\langle{\downarrow}|Q_{33}|{\downarrow}\rangle&=\langle0|Q_{33}|0\rangle
-{2\over {3M}} \langle0|J_{3}|0\rangle, \qquad
\langle{\downarrow}|Q_{33}|{\uparrow}\rangle=
-{1\over3M}\langle0|J_+|0\rangle.\cr}
\numbereq\name{\eqnxzui}
$$

To derive the sum rules for the physical states of the superspin-$j$
multiplet, we add the supersymmetry generated spin ($|{\uparrow}\rangle$
or $|{\downarrow}\rangle$) to the original spin using the appropriate
Clebsch-Gordan combinations
$$
\eqalign{
|j+\half , m+\half\rangle&
={1\over {{\sqrt{2j+1}}}}
\big[ {\sqrt{j+m+1}}|j,m;\uparrow\rangle
+{\sqrt{j-m}}|j,m+1;\downarrow\rangle \big], \cr
|j-\half, m+\half\rangle&={1\over {{\sqrt{2j+1}}}}
\big[ -{\sqrt{j-m}}|j,m;\uparrow\rangle
+{\sqrt{j+m+1}}|j,m+1;\downarrow \rangle \big]. \cr}
\numbereq\name{\eqivic}
$$
For a given state of spin $j$, the quadrupole moment is characterized by
a single quantity, $Q^{(j)}$.  By convention this is taken to be the
matrix element of $Q_{33}$ in the $|j,m=j\rangle$ state.  More
generally, by the Wigner-Eckart theorem, we may write ($j\geq 1$)
$$
Q^{(j)}={j(2j-1)\over3m^2-j(j+1)}
\langle\alpha, j, m| Q_{33} |\alpha,j, m\rangle.
\numbereq\name{\eqkartelo}
$$
We now recall that the massive superspin-$j$ supermultiplet contains
four states, of spins $(j+\ha,j,j,j-\ha)$.  Using the vanishing sum rule
and the ability to choose $|{\updownarrow}\rangle$ as an alternate
Clifford vacuum [\por], we note that both spin-$j$ states have the same
gravitational quadrupole moment.  Thus for the four states there are
three separate quantities, $(Q^{(j+\ha)},Q^{(j)},Q^{(j-\ha)})$, which are
related by the sum rules ($j\geq 3/2$):
$$
Q^{(j+{1\over 2})}=Q^{(j)}+{{2j}\over {3M}}, \qquad
Q^{(j-{1\over 2})}=Q^{(j)}{{(j-1)(2j+3)}\over {j(2j+1)}}
-{2\over 3M}{{(j-1)(2j-1)}\over {(2j+1)}}.
\numbereq\name{\eqzahovic}
$$
For $j=1/2$ and $j=1$ the sum rules simplify to
$$
Q^{(1)}={1\over 3M}\qquad \left(j={1\over 2}\right),\qquad
Q^{({3\over 2})}=Q^{(1)} + {2\over 3M}\qquad (j=1). \numbereq\name{\onehalf}
$$
The gravitational quadrupole moment, $Q_{ij}$, may be related to an
$h$-factor, which is the gravitational analog of the electromagnetic
$g$-factor [\giamn,%
\refx\kri{I. B. Khriplovich, \spn 69 (1989) 217.},%
\refx\krpo{I. B. Khriplovich and A. A. Pomeransky, \spn 86 (1998) 839,
{\tt gr-qc/9710098}.},%
\refx\krpt{I. B. Khriplovich and A. A. Pomeransky, {\tt gr-qc/9809069}.}].
The $h$-factor may be defined in terms of the coupling of a spin-$j$
field $\varphi$ to the Riemann tensor that arises in the equations of
motion as
$$
(\nabla_\mu\nabla^\mu-M^2)\varphi
+h[R_{\mu\nu\lambda\sigma}\half\Sigma^{\mu\nu}\half\Sigma^{\lambda\sigma}]
\varphi+\cdots=0,
\numbereq\name{\eqawer}
$$
where $\Sigma^{\mu\nu}$ is the Lorentz generator in the spin-$j$
representation.  In the non-relativistic limit the Riemann coupling in
(\eqawer) gives rise to a shift in energy
$$
\Delta E=-{h\over 2M} [R_{ijlk}\half\Sigma^{ij}\half\Sigma^{kl}]
= {h\over 2M}J^iJ^j\partial_i\partial_j\phi,
\numbereq\name{\eqnew}
$$
where we have used
the fact that in four dimensions the components of the
Riemann tensor are given in terms of the Newtonian potential
$\phi$ as
$$
R_{ijkl}=
\delta_{ik}\partial_j\partial_l\phi+\delta_{jl}\partial_i\partial_k\phi
-\delta_{il}\partial_j\partial_k\phi-\delta_{jk}\partial_i\partial_l\phi.
\numbereq\name{\eqvboc}
$$
Note that, out of the original Lorentz generators, only the angular
momentum generators $\Sigma^{ij}=\epsilon^{ijk}J_{k}$ remain in this limit.

The $h$-factor can now be related to the gravitational quadrupole
$Q_{ij}$ by recalling that the change in energy of a particle in a static,
slowly varying external Newtonian potential $\phi$ due to the
quadrupole is
$$
\Delta E={1\over 2}Q^{ij}\partial_i\partial_j\phi.
\numbereq\name{\eqenerg}
$$
Comparing (\eqnew) with (\eqenerg), we obtain the relation
$$
Q_{ij}={h\over M}(J_{i}J_{j}-{1\over 3}{\delta_{ij}}J^2),
\numbereq\name{\eqvcouhjl}
$$
where the latter term is necessary to make $Q_{ij}$ traceless.
Nevertheless this has no effect in Eq.~(\eqenerg) since
$\partial_i\partial_i\phi=0$ in vacuum.  Note the resemblance between
the relations (\eqvcouhjl) for the $h$-factor and $\mu_i={ge\over2M}J_i$
for the $g$-factor.

Combining the $h$-factor definition, (\eqvcouhjl), with the quadrupole
matrix element, (\eqkartelo), we find $h_j = {3M\over j(2j-1)}Q^{(j)}$
(where the quadrupole moment is only well defined for spins
$\ge1$).  In terms of $h$-factors, the quadrupole sum rules,
(\eqzahovic), now take on the final simpler form:
$$
h_{j+{1\over 2}}=h_j-{{2(h_j-1)}\over {2j+1}}, \qquad h_{j-{1\over 2}}
=h_j+{{2(h_j-1)}\over {2j+1}} \qquad \left(j\geq {3\over 2}\right),
\numbereq\name{\eqtzovani}
$$

$$
h_1=1\qquad \left(j={1\over 2}\right),
\qquad h_{{3\over 2}}={1\over 3}(h_1+2) \qquad (j=1).
\numbereq\name{\specialcase}
$$
The terms proportional to $(h_j-1)$ are related to a gravitational transition
moment between the $j\pm\ha$ states of the multiplet, and vanish when
$h=1$, yielding the possible value of $h=1$ for all members of the
superspin-$j$ multiplet.  This result from supersymmetry matches with
those obtained previously [\cpd,\krpo,\krpt,\giamn] that similarly
suggest that $h=1$ is a natural value for the $h$-factor, in much the
same way that $g=2$ is natural for the gyromagnetic ratio.

Finally we would like to extend our results to theories invariant
under $N=1$ supersymmetry but violating $R$ symmetry. In this case
the resulting multiplet of currents contains $(j_{\mu}^{(5)}, j_{\mu},
T_{\mu\nu}, P, Q)$, with the following transformation properties:
$$
\eqalign{
{\delta}j_{\mu}^{(5)}&=i {\overline{\epsilon}}{\gamma_5}j_{\mu}
-{i\over 3}{\overline{\epsilon}}{\gamma_5}{\gamma_{\mu}}{\gamma^{\rho}}
j_{\rho},\cr
{\delta}j_{\mu}&=[-2i{\gamma^{\lambda}}T_{\mu\lambda}-{\gamma_5}
{\gamma^{\lambda}}{\partial_{\lambda}}j_{\mu}^{(5)}+{\gamma_5}
{\gamma_{\mu}}
{\partial^{\lambda}}j_{\lambda}^{(5)}
-{i\over 2}{\epsilon_{\mu\nu\lambda\sigma}}
{\gamma^{\nu}}{\partial^{\lambda}}j^{\sigma (5)}+{1\over 3}{\gamma_{\mu\nu}}
{\partial^{\nu}}(P+i{\gamma_5}Q)]{\epsilon},\cr
{\delta}T_{\mu\nu}&=-{1\over 2}\overline{\epsilon}\gamma_{(\mu}{}^\lambda
{\partial_{\lambda}}j_{\nu)},\cr
{\delta}P&=-i{\overline{\epsilon}}{\gamma^{\rho}}
j_{\rho}, \qquad {\delta}Q={\overline{\epsilon}}
{\gamma_5}{\gamma^{\rho}}j_{\rho}.\cr}
\numbereq\name{\eqocnv}
$$
It follows then that two successive supersymmetry transformations on the
conserved current $T_{\mu\nu}$ gives
$$
\eqalign{
{\delta}_{\eta}{\delta}_{\epsilon}T_{\mu\nu}=&
i{\overline{\epsilon}}\gamma_{(\mu}{}^{\lambda\rho}\eta
{\partial_\lambda}T_{\nu)\rho}-i{\overline{\epsilon}}{\gamma^{\lambda}}\eta
{\partial_\lambda}T_{\mu\nu}
+{3\over 4}\overline{\epsilon}\gamma_5\gamma^{\lambda}\eta
(\Box\eta_{\lambda(\mu}^{\vphantom{(5)}}
-\partial_{\lambda\vphantom{)}}^{\vphantom{(5)}}
\partial_{(\mu}^{\vphantom{(5)}}) j_{\nu)}^{(5)}\cr
&-{1\over 4}\overline{\epsilon}\gamma_5\gamma^{\lambda}\eta
(\Box\eta_{\mu\nu} -\partial_\mu\partial_\nu) j_{\lambda}^{(5)}
+{3\over 4}\overline{\epsilon}\gamma_5\gamma_\lambda\eta
(\eta_{\mu\nu}\eta^{\lambda\sigma}-\delta_{(\mu}^\lambda\delta_{\nu)}^\sigma)
\partial_\sigma\partial^\rho j_{\rho}^{(5)}\cr
&+{1\over6}\overline{\epsilon}\eta
(\Box\eta_{\mu\nu}-\partial_\mu\partial_\nu)P
+{i\over6}\overline{\epsilon}\gamma_5\eta
(\Box\eta_{\mu\nu}-\partial_\mu\partial_\nu)Q . \cr}
\numbereq\name{\eqpavxc}
$$
While added terms arise due to the non-conservation of $j_\mu^{(5)}$,
they nevertheless do not contribute to the double variation of
${\cal T}_{00}$ to the order that would affect the quadrupole matrix
elements.  Thus $\delta_{\eta_R}\delta_{\epsilon_L}{\cal T}_{00}$ is
unchanged from (\eqddct), and hence the resulting sum rules for the
gravitational quadrupole moments coincide with those derived above for
the $R$-invariant theory.

Although we have focused our attention on gravitational quadrupole
moments, the sum rules we have derived may easily be generalized along
the lines of [\giamn] to encompass arbitrary gravitational multipoles.
While such higher moments may lack clear physical significance, they
many nevertheless provide a means for examining higher derivative
couplings to gravity, especially in a string context.  Additionally,
the extension of the above to the case of extended supersymmetry
should allow for the study of graviphoton sum rules in a model
independent manner since the corresponding currents would indeed fall
into the gravity sector of the theory.

\vskip .1in
\noindent
{\bf Acknowledgments.} \vskip .01in \noindent
M.P. would like to thank the ITP at UCSB for its kind hospitality during 
completion of this work.
This work was supported in part by the Department of Energy under Contract
Number DE-FG02-91ER40651-TASK B, and by the NSF under grant PHY-9722083.

\immediate\closeout1
\bigbreak\bigskip

\line{\twelvebf References. \hfil}
\nobreak\medskip\vskip\parskip

\input refs

\vfil\end

\bye